\documentclass[jap,aip,10pt,twocolumn,superscriptaddress,showpacs]{revtex4-1}

\usepackage{graphics}
\usepackage{natbib}

\begin{document}

\title{Physical limitations in ferromagnetic inductively coupled plasma sources}

\author{Yury P. Bliokh}
\affiliation{Department of Physics, Technion-Israel Institute of Technology, Haifa 32000,
Israel}

\author{Joshua Felsteiner}
\affiliation{Department of Physics, Technion-Israel Institute of Technology, Haifa 32000,
Israel}

\author{Yakov Z. Slutsker}
\affiliation{Department of Physics, Technion-Israel Institute of Technology, Haifa 32000,
Israel}

\begin{abstract}
The Ferromagnetic Inductively Coupled Plasma (FICP) source, which is a version of the
common inductively coupled plasma sources, has a number of well known advantages such as
high efficiency, high level of ionization, low minimal gas pressure, very low required
driver frequency, and even a possibility to be driven by single current pulses. We
present an experimental study of such an FICP source which showed that above a certain
value of the driving pulse power the properties of this device changed rather
drastically. Namely, the plasma became non-stationary and non-uniform contrary to the
stationary and uniform plasmas typical for this kind of plasma sources. In this case the
plasma appeared as a narrow dense spike which was short compared to the driving pulse.
The local plasma density could exceed the neutral atoms density by a few orders of
magnitude. When that happened, the afterglow plasma decay time after the end of the pulse
was long compared to an ordinary case with no plasma spike. Experiments were performed
with various gases and in a wide range of pressures which enabled  us to understand the
physical mechanism and derive the parameters responsible for such plasma behavior. A
qualitative model of this phenomenon is discussed.
\end{abstract}


\maketitle

\section{Introduction}

Inductively Coupled Plasma (ICP) sources are known to have been widely implemented.
Indeed these electrode-less, often rf driven sources require neither high voltages nor
high currents. Typically they require a few hundreds of volts at a few tens of amperes,
and the used frequency as a rule is below 15 MHz \cite{1}.  These plasma sources are
capable of producing dense low-noise uniform plasmas in a large volume. The plasma
density might vary within a wide range: $10^9$-$10^{12}$ cm$^{-3}$ at electron
temperature which could be as low as 1 eV \cite{2}.  The plasma size may achieve several
tens of cm. Such sources are widely used for many purposes, such as plasma processing
(etching, surface cleaning, sputtering), light sources, etc. \cite{1, 3}.  A distinct
family of these ICP sources is Ferromagnetic ICP (FICP) sources. In an ordinary ICP
device, plasma exists due to the induced electric field near the inductor-like antenna
when rf voltage is applied to its input. The electric field is maximal in the inductor
neighborhood and falls down to zero in the center, while the rf magnetic field of the
inductor fills the whole volume. In the FICP the antenna appears as a magnetic core with
a few turns of winding, which is fully immersed in the plasma \cite{4, 5}.  In this case
the electric field fills the whole device and does not vanish at the core axis while most
of the magnetic field of the winding is inside the magnetic core and does not penetrate
into the plasma. Due to the magnetic core, the inductance of the FICP is high. Therefore,
it is possible to drive this plasma source with comparatively low frequency and even to
work with a relatively long single pulse which significantly simplifies the driver. The
FICP device may in general achieve considerably denser plasma due to the greatly
increased coupling brought on with the closed magnetic core.

In practice, there are two versions of such FICP device. The first one appears as a
single comparatively thin toroidal ferromagnetic core having a large diameter \cite{4}.
The opening of this core should exceed the required plasma size. In the second version
the FICP consists of a large number (few tens) of small ferromagnetic cores which are
properly connected to each other \cite{5}. These cores may be placed in space in various
ways: in one plane, on a cylinder surface, etc. Both of these versions have the same
advantages, such as very high efficiency (up to 99\%) and very low minimal working
pressure (about $10^{-4}$ Torr). These parameters are definitely better as compared to
ordinary coreless ICP. The FICP devices, being driven by rf current \cite{5, 6}, require
a low frequency, typically 240-280 kHz, and with this frequency they show a very high
$\cos\varphi>0.9$. The latter means almost pure active loading of the plasma source
driver (e.g., an rf oscillator) which is a significant advantage over coreless ICP. For
both these versions the driver fitting is very simple: just the number of turns winding
should be chosen correctly like in ordinary transformers. The differences between these
two FICP versions are the following: the multi-core version \cite{5} is able to produce a
more uniform plasma (the non-uniformity could be below 3-4\%) as well as it is able to
form a required plasma density gradient, but the maximal plasma density usually did not
exceed $2\cdot10^{12}{\rm cm}^{-3}$. The single-core FICP \cite{4, 7}  forms less uniform
plasmas (the non-uniformity is about 7-10\%) for the same plasma size (20-30 cm), and
there is no option to vary the spatial distribution of the produced plasmas, but it is
able to produce denser plasmas, up to $10^{13} {\rm cm}^{-3}$, with ionization rate above
90\%. Another advantage of the single-core device is that its input resistance is often
independent of the plasma density \cite{8}.

As we have already mentioned, due to the high initial inductance of the FICP device, it
is possible to drive it even by a single pulse, e.g. by discharging a preliminary charged
capacitor via the FICP primary winding \cite{4}. However, almost all studies of this
device were performed with various rf drivers \cite{5, 6, 7, 8, 9} because of a clear
reason: in this case the FICP devices produce stationary plasmas where the plasma exists
during the whole operation time of the rf oscillator, including CW regime. On the other
hand these powerful (10-15 kW) oscillators could be rather sophisticated and expensive
units. The only single-pulse ``exception'' was our early work \cite{4}, but even in that
work a ballast inductor was connected in series to the primary winding. This was done in
order to restrict the primary current, i.e., to eliminate the influence of the produced
plasma on the pulse parameters and to expand the pulse. It is obvious that in this case
processes of plasma creation and charged particles losses, as well as spatial and
temporal plasma evolution, were ``externally'' disturbed. Advantages of single-pulse
drivers are simplicity, low cost and, most importantly, very high pulsed power.

In this paper we present results of experimental study of an FICP plasma source with a
single-pulse driver with no ballast and whose output impedance was minimized. Compared to
the former experiments \cite{4}, we significantly increased the voltage across the plasma
(by a few times) and the current through the plasma (by an order of magnitude), no matter
whether they were rf-driven or operated in the single-pulse mode. The power delivered to
the FICP could significantly increase the values needed for 100\% ionization. Under these
conditions the plasma evolution was investigated. This self consistent evolution resulted
in drastic changes of the obtained plasma parameters, which significantly differ from
those obtained for low power. This, in turn, shows that there is no sense to raise the
driving power above a certain limit. These results could be important for optimal design
of FICP sources.

\section{Apparatus description}

\begin{figure}[htb]
\centering \scalebox{0.4}{\includegraphics{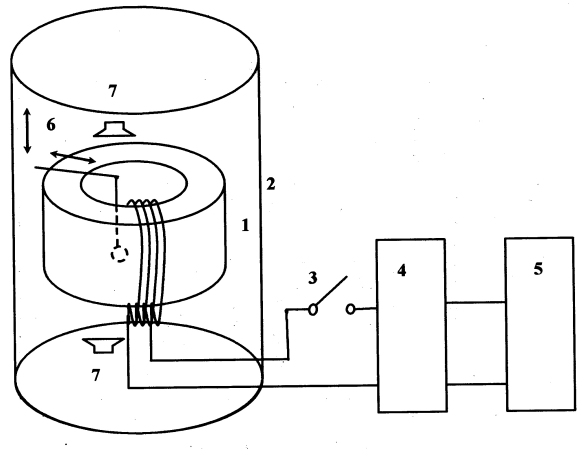}}
\caption{Experimental setup: (1) Supermendure magnetic core;  (2) glass vacuum vessel;
(3) electronic key;  (4) storage capacitor, voltage divider and current transformer;  (5)
rectifier;  (6) movable probe;  (7) set of microwave antennas. \label{Fig_1}}
\end{figure}

The experiments were carried out in a glass vacuum vessel of 32 cm diameter and 50 cm
height, similar to our earlier work \cite{4, 9}  (see Fig.~\ref{Fig_1}). This vessel was pumped to
pressure $p$ of about $2\cdot10^{-4}$ Torr and then it was filled with He, Ar, or Xe. The
measurements were performed within a pressure range of $10^{-4}-2.5\cdot10^{-2}$ Torr. As
a ferromagnetic core we used a Supermendure core having 15 cm outer diameter, 10 cm inner
diameter and 5 cm height. The Supermendure core was chosen because of its high saturation
level of about 3 T. It was surrounded by three coils, two of which consisted of 10 turns
of winding and the third one consisted of just one turn. One of the 10-turn coils (shown
in Fig. 1) was used as a primary coil and was connected via an electronic key to the
charged capacitor. This 1 $\mu$F capacitor could be charged up to 1.5 kV. The primary
coil was shunted by a high-current diode to prevent the influence of the driver on the
primary current after delivery of the stored energy to the plasma. Such scheme does not
perturb the current caused by the plasma self inductance. The second 10-turn coil was
used to eliminate the magnetization of the core. We passed through it a 2 A dc current
which was sufficient. To prevent the influence of the dc-current source, we used a chock
of about 100 mH inductance, which was sufficient as well. The third coil, namely the
1-turn coil, was used for diagnostics: to measure the voltage per turn and to control the
absence of core saturation.

To measure the current through the primary winding (primary current $I_{\rm pr}$) we used
either a Current View Resistor (CVR) of $0.1\Omega$ or a Current Transformer (CT), the
discrepancy between them never exceeded a very few percents. A thin (about 2 cm) CT
having a large (13 cm) diameter was designed, built and calibrated. This CT could be
placed directly at the FICP top and it was used to measure the total current $I_{\rm pl}$
induced in the discharge plasma. When  the $I_{\rm pr}$  and  $I_{\rm pl}$  waveforms
were similar and the transformer ratio  $I_{\rm pl}/I_{\rm pr}$  was close to 10, i.e.,
the number of the primary turns, there was no core saturation and the core losses were
minimal. A very small discrepancy might be just at the beginning of the driving (primary)
current $I_{\rm pl}$ during 0.5-1.5 $\mu$s -- the time of the gas breakdown, which
increased with the pressure reduction. When there was no discharge there was no
demagnetizing current through the plasma and as a consequence the primary current
waveform and value became very different. This indicated a minimal pressure p for each
sort of gas, namely 0.1 mTorr for Xe,  0.3 mTorr for Ar, and 5 mTorr for He. In all these
cases the measurements were started from plasma current $I_{\rm pl}=200-300$A, which was
actually the maximal current for all former experiments \cite{4,5,6,7,8,9}.

In the presented experiments we used a low-power long-pulse hot-cathode discharge for
plasma ignition. It is not shown in Fig.~\ref{Fig_1}, but it was described in detail in our recent
work \cite{10}. For plasma diagnostics we used combination of two methods: the plasma
probing and microwave cut-off methods \cite{4,6,7,8,9,10}. To measure the parameters of
the dense plasma produced by FICP during a high-current pulse we used a small
semi-spherical single probe with collecting area of about 5 mm$^2$. This probe was well
shielded, its outer diameter was about 4 mm. This probe mainly worked in the ion
saturation current regime, so the area of its holder (4 mm diameter, 100 mm length) was
quite sufficient to be used as a base electrode. To prevent parasitic signals this holder
as well as the probe bias supply were insulated from the ground. To measure the probe
current we also used a small current transformer. This probe was radially and axially
movable. For microwave cut-off diagnostics we used 3 sets of transmitters and detectors.
One set was for 9.5 GHz (critical plasma density $n_c=6\cdot 10^{12}{\rm cm}^{-3}$),
another one for 22.5 GHz ($n_c=7\cdot 10^{12}{\rm cm}^{-3}$) and the third one for 70 GHz
($n_c=6\cdot 10^{13}{\rm cm}^{-3}$). The transmitting antennas were placed above the
upper FICP opening and the receiving antennas were placed below the bottom FICP opening,
all of them were close to the FICP center.

\section{Experimental results}

\begin{figure}[htb]
\centering \scalebox{0.5}{\includegraphics{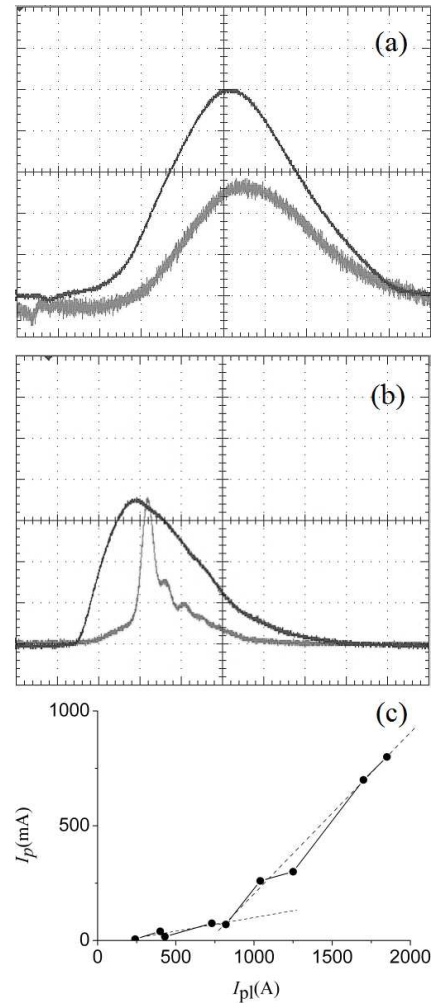}}
\caption{ a -  waveforms of the plasma current $I_{\rm pl}$ (upper trace, 50 A/div) and of the ion
current $I_p$ collected by probe (lower trace, 2 mA/div) for low $I_{\rm pl} = 250$ A;  b -  the
same for high $I_{\rm pl}  = 1700$ A (upper trace, 500 A/div), lower trace for $I_p$, 200 mA/div;  c
-  dependence of $I_p$ on  $I_{\rm pl}$. All data for $p = 1$ mTorr of Ar. \label{Fig_2}}
\end{figure}

When the negative bias applied to the probe exceeded a certain value (35 V for Xe, 45 V
for Ar and 70 V for He) the ion probe current $I_p$ reached saturation. As a rule we
worked with probe bias above these values. If the current induced in the plasma $I_{\rm
pl}$ during the discharge pulse did not exceed a certain value (200-400 A), the waveform
of probe current $I_p$  was similar to the waveforms of the primary winding current
$I_{\rm pr}$ and to the induced current $I_{\rm pl}$ (Fig.~\ref{Fig_2}a). In the experiments
described in Ref.~\cite{4}  the plasma current never exceeded this value. However, with
the increase of  $I_{\rm pl}$  the $I_p$ waveform changed gradually. Above a certain
threshold value of $I_{\rm pl}$ the maximal value of $I_p$ increased significantly and
its waveform became narrow and sharp (Fig.~\ref{Fig_2}b). The dependence of the  $I_p$  maximum on
$I_{\rm pl}$ is shown in Fig.~\ref{Fig_2}c. It is seen there that the maximal value of the probe
current  $I_p$ increases abruptly starting from a certain threshold value $I_{\rm th}$ of
the plasma current $I_{\rm pl}$. As it is seen  in Fig.~\ref{Fig_2}c, for $p=1$ mTorr of Ar,
$I_{\rm th}=950$ A. Very similar results were obtained with He and Xe, just the values of
the maximal probe current $I_p$ and plasma threshold current $I_{\rm th}$ were different for
various gases and pressures. For each gas the threshold current $I_{\rm th}$ increased
slowly but monotonically with the increase of  $p$:  from 1050 A  at $p=0.1$ mTorr  to
2050 A at $p=1$ mTorr for Xe,  and from 650 A  at  5 mTorr  to 1150 A at 25 mTorr for He.

\begin{figure}[htb]
\centering \scalebox{0.72}{\includegraphics{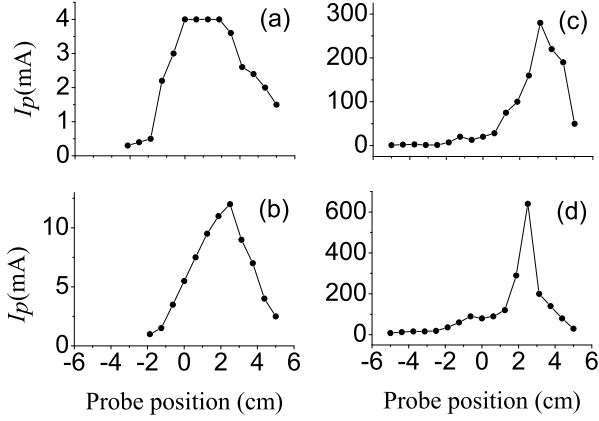}}
\caption{Spatial distribution of ion probe current in the middle cross sectional area of
the magnetic core for Xe at various pressures $p$ and plasma currents $I_{\rm pl}$.  a-  $p = 0.1$
mTorr, $I_{\rm pl} = 430$ A; b-  $p = 1$ mTorr, $I_{\rm pl} = 480$ A; c-  $p = 0.1$ mTorr, $I_{\rm pl} = 1100$ A;  d-
$p = 1$ mTorr, $I_{\rm pl} = 2700$ A\label{Fig_3}}
\end{figure}

\begin{figure}[htb]
\centering \scalebox{1.55}{\includegraphics{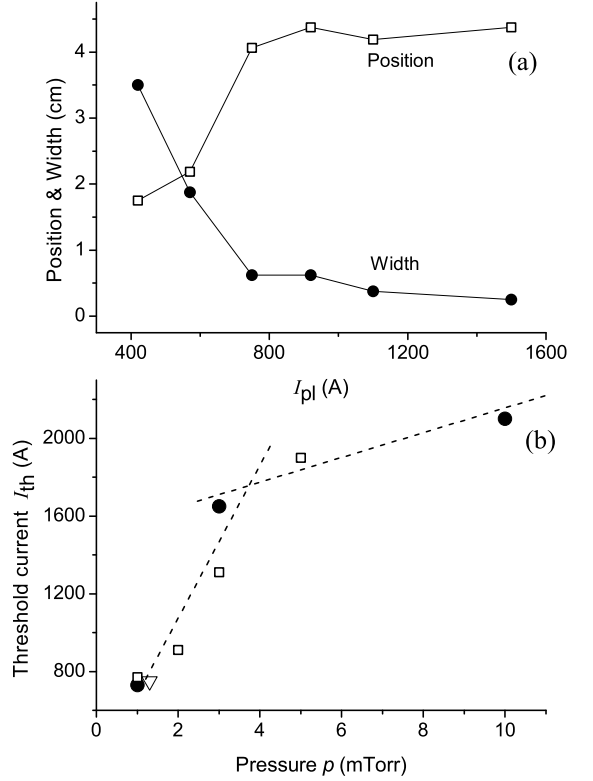}}
\caption{a -  location of maximal $I_p$ \textit{vs} plasma current $I_{\rm pl}$ (upper curve) and its width at 70\% of its maximum;  b -  plasma compression threshold current \textit{vs} pressure $p$ for He (circles), Ar (triangle) and Xe (squares). Here the true pressure for Xe has been multiplied by 10 and for He it has been divided by 5.\label{Fig_4}}
\end{figure}

The increase of the plasma current $I_{\rm pl}$ caused not only the decrease of the probe current
$I_p$ but also its strong dependence on the probe position. Two examples of this dependence
in the middle cross sectional area of the FICP core are shown in Fig.~\ref{Fig_3}a,b in the case
when $I_{\rm pl}<I_{\rm th}$. They are comparatively smooth and very similar to those obtained in
Refs. \cite{4,7}. Above the threshold, when  $I_{\rm pl}>I_{\rm th}$, a narrow ``filament'' of $I_p$ appears
in parallel to the core axis somewhat off center (Fig.~\ref{Fig_3}c,d).
The probe currents $I_p$ in
the core openings (top and bottom) were approximately the same and just 20-25\% less
than in the middle core cross sectional area. It should be noted that the resolution of
these measurements could not be better than the shielded probe diameter, i.e. 4 mm. A
typical evolution of the filament location and its width \textit{vs} the plasma current $I_{\rm pl}$ is shown in Fig.~\ref{Fig_4}a. The smallest measured filament width was never less than 5 mm, probably because of  the probe resolution. Although for various gases and pressures there were different values of $I_{\rm th}$, the location of the $I_p$ maximum was the same. It is also interesting
to note that multiplying the measured pressure of Xe by 10 and dividing the measured
pressure of He by 5, all dependencies of $I_{\rm th}$ on $p$ for He, Ar, and Xe may be put together on
the same curve, as it is shown in Fig.~\ref{Fig_4}b. These factors of 10 and 5 are close to the
ratios of the corresponding ionization cross sections within an order of magnitude.

\begin{figure}[htb]
\centering \scalebox{0.4}{\includegraphics{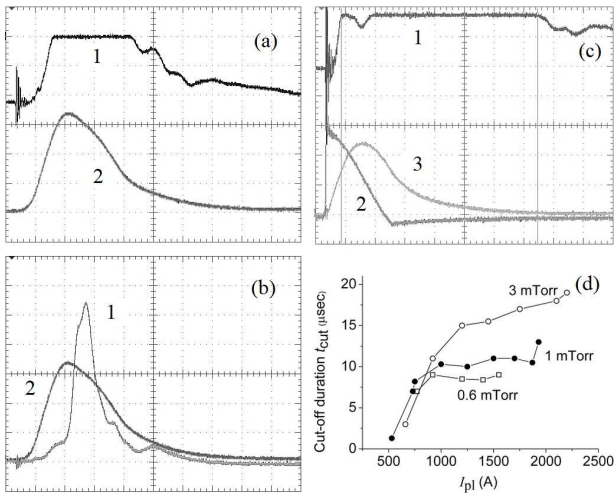}}
\caption{ a -  70 GHz microwave cut-off (trace 1, 20 mV/div, negative signal with zero
level at third horizontal division) and plasma current $I_{\rm pl}$ (trace 2, 500 A/div);  b -
probe current $I_p$ (trace 1, 50 mA/div) and plasma current $I_{\rm pl}$ (trace 2, same as in a);
pressure $p = 1$ mTorr of Ar for both a and b;   c -  maximal duration of 70 GHz microwave
cut-off, obtained with Xe at maximal pressure (1 mTorr) and maximal plasma current (2600
A).  Trace 1 corresponds to the microwave cut-off (20 mV/div, negative signal with zero
level at fourth horizontal division), trace 2 corresponds to the voltage across the
primary winding (500 V/div) and trace 3 to the plasma current $I_{\rm pl}$ (1000 A/div);  d - the
microwave cut-off duration $t_{\rm cut}$ \textit{vs} the plasma current $I_{\rm pl}$ for various pressures of Ar.\label{Fig_5}}
\end{figure}

To
connect the measured ion probe currents to the plasma densities we used the microwave
cut-off method (see e.g. \cite{6,7,8,9} and refs. therein). Scope traces for a typical microwave
cut-off signal for 70 GHz (critical density $n_c=6\cdot10^{13}{\rm cm}^{-3}$) and the plasma current $I_{\rm pl}$ are shown
in Fig.~\ref{Fig_5}a for Ar at $p=1$ mTorr. The ion probe current $I_p$ and the plasma current $I_{\rm pl}$ for the
same case are shown in Fig.~\ref{Fig_5}b. Similar waveforms were also obtained for He and Xe. For
all sorts of gas and pressures the microwave cut-off appeared if the plasma current $I_{\rm pl}$
was high enough and the cut-off existed just within a certain range of this plasma
current. Strictly speaking, at the moments when the cut-off starts and ends, the
corresponding plasma currents might not be equal to each other. Despite the fact that the
maximal $I_{\rm pl}$ could exceed 2.5 kA, these currents could usually be found within the range
of 600-1400 A for Ar and 300-700 A for Xe. Typically $I_{\rm pl}$ is larger in the beginning.
These plasma currents tend to be smaller at higher pressures and for heavy gas. For Xe at $p=1$
mTorr  and $I_{\rm pl}=2.6$ kA, the minimal cut-off holding current $I_{\rm pl}$ is about 100 A (Fig.~\ref{Fig_5}c). Note, that at the end of the pulse the plasma current exists due to the plasma self
inductance, which is clear from the waveform of the voltage across the primary winding
(Fig.~\ref{Fig_5}c). Indeed, this voltage changes polarity during the pulse while the plasma
current does not. When the cut-off appeared with the heavier gases, Ar and Xe, the value
of the probe current $I_p$ at the cut-off moment depended on the sort of gas but was almost
independent of $p$ and $I_{\rm pl}$.  For Ar  $I_p=49\pm2$ mA (averaged over 16 measurements) and
for Xe   $I_p=23\pm 1.6$ mA (averaged over 10 measurements). Also, when the cut-off
appeared, its duration $t_{\rm cut}$ increased very fast with the increase of $I_{\rm pl}$ and then
saturated as it is seen in Fig.~\ref{Fig_5}c. It is also seen there that if the plasma current is
kept constant ($I_{\rm pl}={\rm const}$), the cut-off duration $t_{\rm cut}$ tends to become higher with the
increase of $p$:  for Ar in the pressure range $p=0.6 – 3$ mTorr,  $t_{\rm cut}$ increased from 9 $\mu$s to
18 $\mu$s. Also $t_{\rm cut}$ increased with the atomic weight of the gas. However, we never obtained
$t_{\rm cut}$ above 30 $\mu$s (for Xe at $p=1$ mTorr and maximal $I_{\rm pl}=2.6$ kA). This corresponds to the
mentioned above minimal $I_{\rm pl}$ where the cut-off stops at the end of the current pulse
(Fig.~\ref{Fig_5}c).

The results obtained with the lightest gas we used (He) were qualitatively
similar but with one exception. Namely, when $I_{\rm pl}$ reached the value needed for microwave
cut-off, even a small overlap was enough for $t_{\rm cut}$ to saturate. When it happened, the
probe current $I_p$ in the cut-off beginning differed from $I_p$ at the cut-off end. Thus, the
cut-off started with probe current  $I_p=120\pm8$ mA and ended with $I_p=250\pm 17$ mA. This
result is correct for the whole pressure and plasma current ranges, 5-25 mTorr and 200-2000 A, respectively. At the minimal pressure for each sort of gas the collected probe
current $I_p$ became unstable and the cut-off signal waveform became irregular.

These
measurements were carried out with a 70 GHz microwave signal. The results allows one to
estimate the effective electron temperature $T_{\rm eff}$ at the moments when cut-off starts and ends.
Indeed, at these moments the plasma density is equal to the critical density $n_c=6\cdot 10^{13}{\rm cm}^{-3}$. For
such plasma density and even for unrealistic high electron temperature of 70 eV (the
highest bias voltage to get $I_p$ saturation), the Debye length should not exceed $10^{-2}$ mm,
i.e. much less than the probe sizes, that allows one to consider the probe as planar. For
planar probes, there is a simple relation between the ion saturation current $I_p$ and
electron temperature $T_e$ (for Maxwellian plasmas): $T_e=MI_p^2/(AenS)^2$, where $A=0.4$ is a dimensionless
coefficient \cite{9}, $M$ is the ion mass and $S$ is the probe area ($S=5 {\rm mm}^2$ in the
experiments). Taking into account that in FICP sources the electron distribution function
is not exactly Maxwellian (see \cite{9} and refs. therein) we regarded the derived result as a
certain effective electron temperature $T_{\rm eff}$. So, we obtained $T_{\rm eff}=1.5$ eV for Xe, $T_{\rm eff}=2.5$ eV for
Ar, and  $T_{\rm eff}=1.5$ eV and 6 eV at the beginning and end of the cut-off for He. The routine
usage of the probe current-voltage characteristics to derive the electron temperature was
not relevant in our case. When the probe bias was below a few tens of volts, as needed
for ion current saturation, the probe current became affected by high energy electron
beams, related to the voltage induced in the plasma and typically present in such plasma
sources \cite{9}. As a result, the probe current became very sensitive to the voltage falls
and currents in the plasma, i.e. to the regime of the plasma source. So, derivation of
the plasma parameters from probe characteristics became problematic, in particular taking
into account that in the present experiments the induced voltages and currents are
definitely higher than even those in Ref. \cite{8}. Oppositely, when the probe bias is
sufficiently high, the ion saturation current $I_p$ is proportional to the plasma density $n$
and almost insensitive to the above mentioned factors \cite{9,11}. Comparing the voltage
required to get the saturation and the evaluated values of $T_{\rm eff}$ (tens of volt \textit{vs} volts), one
might expect that intensive electron beams exist in the ``cold'' bulk plasma. The evaluated
$T_{\rm eff}$ was about 2-3 times smaller than the one obtained formerly in these plasma sources \cite{4,5,6,7,8,9}. To verify this point we compared the decay time of the afterglow plasma $t_{\rm dec}$ at high and
low plasma current $I_{\rm pl}$. To do this, we used microwave oscillators for 9.5 GHz ($n_c=1.2\cdot10^{12} {\rm cm}^{-3}$) and
for 24.6 GHz ($n_c=7.7\cdot10^{12} {\rm cm}^{-3}$) because for low $I_{\rm pl}$ the plasma density is definitely smaller. Thus,
with $p=3$ mTorr of Ar and $I_{\rm pl}=300$ A, we obtained the microwave cut-off at 24.6 GHz a
bit after the $I_{\rm pl}$ pulse and the cut-off at 9.5 GHz at about 250 $\mu$s later. Taking into
account the corresponding densities ratio (about 6 times), one might easily derive the
plasma decay time $t_{\rm dec}=120 \mu$s. The same measurements done with $I_{\rm pl}=1800$ A showed for the
microwave frequency of 24.6 GHz a cut-off delay of about 300 $\mu$s and for the 9.5 GHz --
about 800 $\mu$s. The difference was about 500 $\mu$s and, consequently, $t_{\rm dec}=270 \mu$s. This decay
time is more than twice longer than in the ``low current'' case, which at least indirectly
confirmed our considerations concerning low $T_{\rm eff}$. This could be qualitatively understood
taking into account that the high-energy fraction of plasma electrons contains more
energy in the ``high current'' case and this fraction disappears very fast during decay
processes, in the absence of induced electric and magnetic fields.

\section{Discussion}

A remarkable result which is clear from the present measurements is the fact that at low
pressures the neutral atoms density $N$ is more than an order of magnitude below the
critical plasma density of $n_c=6\cdot10^{13}\,{\rm cm}^{-3}$  required for microwave
cut-off at 70 GHz. Indeed, for Xe at  $p=0.1$ mTorr  we have $N=3.6\cdot10^{12}\,{\rm
cm}^{-3}$. The reason is the plasma compression – instead of more or less evenly
distributed plasma inside the magnetic core at low $I_{\rm pl}$ \cite{4,5,7}, a thin
cylinder of dense plasma appears somewhat aside of the core axis (Fig.~\ref{Fig_4}a,b ).
Considering the total ionization of the neutral gas inside the core, which was quite
reasonable even with  $I_{\rm pl}=200-300$ A \cite{4,5,6,7,8,9}, it is easy to estimate
the level of such compression. For the minimal pressure of Xe and total ionization, the
density of the uniformly distributed plasma is $n=3.6\cdot 10^{12}\,{\rm cm}^{-3}$.
Taking into account that the probe current is $I_p=23$ mA for Xe and critical plasma
density of $n_c=6\cdot 10^{13}\,{\rm cm}^{-3}$  a probe current of $I_p=1000$ mA (maximal
$I_p$ at $p=0.1$ mTorr) corresponds to a plasma density of $n= 2.5\cdot 10^{15}\,{\rm
cm}^{-3}$. Therefore the compression ratio is about 670. This means that the  initially
uniform plasma for an inner FICP diameter of, say, 10 cm should be reduced by
approximately 26 times (sq. root of 670) to a diameter of about 3.8 mm. This is close to
the minimal size of the plasma that we measured, i.e., close to the spatial resolution of
the probe. It seems likely that plasma compression was caused by the magnetic field,
which, in turn, was caused by the high current $I_{\rm pl}$ induced in the plasma. For
example, suppose that at low gas pressure the plasma compression started from a plasma
current of about 500 A and the corresponding plasma diameter was about 6 cm
(Fig.~\ref{Fig_3}a). Then one could easily derive that the magnetic field $H$ at the
plasma boundary was approximately 30 Gauss. Considering an electron temperature of, say 3
eV,  it is easily seen that the equality  $H^2/8\pi=nT_e$ leads to a plasma density of $n
= 1.5\cdot 10^{13}\,{\rm cm}^{-3}$. This was sufficient for plasma compression because at
low gas pressures (0.1-0.3 mTorr) and total ionization, the plasma density should be
within the range of $3.5\cdot 10^{12}-10^{13}\,{\rm cm}^{-3}$.

A more or less exact description of the plasma dynamics during the comparatively short
pulse is a very sophisticated problem. On one hand, many parameters and processes should
be taken into account, such as ionization and charged particles losses, self consistent
influence of the induced electric fields and currents, etc. On the other hand, it is
quite clear that at low gas pressures just plasma compression may provide a microwave
cut-off for 70 GHz. Note, that at low pressures all the mean free paths of the charged
particles do exceed the core size and just a magnetic field may hold the plasma inside
the device volume, reducing the wall losses. The fact that at least at low pressures the
cut-off starts and ends when the plasma current is quite large, confirms this statement.
On the contrary, when the gas pressure is high, a high ionization rate is not needed to
get the cut-off and the mean free paths are small compared the device size. In this case
collisions should impede the charged particle losses, i.e. there is a certain sense that
the diffusion processes play a role of plasma confinement \cite{7}. As a consequence, a high
plasma current is not needed for the microwave cut-off at high gas pressures which is
also the experimental result:  the microwave cut-off duration $t_{\rm cut}$  increased slowly but
monotonically with the pressure increase (Fig.~\ref{Fig_5}d), and could be comparable to the
duration of the whole current pulse in the case of  Xe (Fig.~\ref{Fig_5}c). In the latter case the
plasma current does not exceed 100 A at the cut-off end.

It is interesting to note, that if the plasma current pulse was strong enough, near its
top the microwave detector received a weak microwave signal for a short time
(Fig.~\ref{Fig_5}c). This is understood because at the highest $I_{\rm pl}$ the
compressed plasma ``string'' might be too thin, even less than the measured 4 mm. Taking
into account that the microwave wavelength was just 4.3 mm and a certain displacement of
the plasma string from the axis took place (Fig.~\ref{Fig_4}a), microwave ``leakage'' was
possible. It should also be noted that at the highest pressures we worked with (5-25
mTorr of He) even the fast fraction of plasma electrons might have enough collisions to
stay longer inside the core \cite{7} and, as a consequence, to further heat the bulk
plasma electrons (in our case from 1.5 to 6 eV).

\section{Conclusion}

We have shown that a number of advantages of FICP sources, such as plasma uniformity,
high ionization rate, absence of magnetic field etc., are naturally limited. These
limitations are not connected to the ferromagnetic core saturation or other
design-related reasons. With the increase  of the plasma driving power, the initially
uniform plasma becomes strongly compressed from the periphery inwards and evolves into a
thin string somewhat near the center. This compression (pinch) might be associated with
the full ionization of neutral gas in the inner core volume. The reason of this plasma
pinch is the self consistent magnetic field which appears in the plasma due to the
induced electric current. This phenomenon is related to the basic principle of this
device operation (actually this is a transformer) and could not be eliminated. This
plasma compression is accompanied with a certain reduction of the bulk plasma electron
temperature.

Consequently, if the aim is a large volume of uniform plasma, the power delivered to the
device should not exceed a certain value, which could be simply estimated. We believe
that this restriction is valid not only for a single-pulse driven FICP but also for
rf-driven versions of such device. On the other hand, if the aim is a long-lasting dense
afterglow plasma, it should be reasonable to exceed this threshold value needed for
plasma compression.

\end{document}